\documentclass[aps,prb,floatfix,twocolumn]{revtex4}
\usepackage{graphicx}
\usepackage{bm}
\usepackage{amsmath}
\usepackage{dcolumn}

\begin{document}

\title{First-principles study of the ferroelastic phase transition in 
$\text{CaCl}_2$}

\author{J.A. V\'algoma}
\author{J.M. Perez-Mato}
\author{Alberto Garc\'{\i}a}
\email{wdpgaara@lg.ehu.es}
\affiliation{
Departamento de F\'{\i}sica de la Materia Condensada, 
Universidad del Pa\'{\i}s Vasco, Apartado 644, 48080 Bilbao, Spain}

\author{K. Schwarz}
\author{P. Blaha}
\affiliation{Institute of Physical and Theoretical Chemistry,
Vienna University of Technology, A-1060 Vienna,
Getreidemarkt 9/156, Austria}

\begin{abstract}
First-principles density-functional calculations within the
local-density approximation and the pseudopotential approach are used
to study and characterize the ferroelastic phase transition in calcium
chloride ($\text{CaCl}_2$). In accord with experiment, the energy
map of $\text{CaCl}_2$ has the typical features of a pseudoproper
ferroelastic with an optical instability as ultimate origin of the
phase transition. This unstable optic mode is close to a pure rigid
unit mode of the framework of chlorine atoms and has a negative
Gr\"uneisen parameter. The ab-initio ground state agrees fairly well
with the experimental low temperature structure extrapolated at
0~K. The calculated energy map around the ground state is interpreted
as an extrapolated Landau free-energy and is successfully used to
explain some of the observed thermal properties.  Higher-order
anharmonic couplings between the strain and the unstable optic mode,
proposed in previous literature as important terms to explain the
soft-phonon temperature behavior, are shown to be irrelevant for this
purpose.  The LAPW method is shown to reproduce the plane-wave results
in $\text{CaCl}_2$ within the precision of the calculations, and is
used to analyze the relative stability of different phases in
$\text{CaCl}_2$ and the chemically similar compound $\text{SrCl}_2$.

\end{abstract}

\maketitle

\section{INTRODUCTION}
In recent years, ab-initio methods based on density-functional theory
have been shown to have predictive power in the field of
thermal structural phase transitions. In particular, first-principles
calculations within the local-density approximation (LDA) have been
able to reproduce ferroelectric and other instabilities in perovskite
oxides such as $\text{BaTiO}_3$, $\text{PbTiO}_3$ or $\text{SrTiO}_3$.
\cite{BaTiO3,PbTiO3,SrTiO3,SrHfO3,BaTiO3LAPW} These ab-initio
calculations correctly predict a low-symmetry distorted ground state
and the eigenvector of the associated distorting mode. Moreover, the
ab-initio exploration of the energy around this ground state for the
relevant degrees of freedom allows, in general, a parametrization of
the obtained energy map as an effective hamiltonian. Thermal effects
investigated through Monte Carlo or Molecular Dynamics simulations
restricted to this effective hamiltonian show a fair agreement with
experiment.

However, the field of temperature-driven ferroelastic phase
transitions, where some spontaneous strain component can be identified
with the order parameter (proper ferroelastic) or is bilinearly
coupled with it (pseudoproper ferroelastic), represents a particular
challenge for first-principles calculations. Density-functional
methods are known to have accuracy problems when
reproducing elastic properties, while in ferroelastics the elastic
energy plays a fundamental role in the structural instability.

We report here the results of an ab-initio study of the ferroelastic
instability in $\text{CaCl}_2$ based on the density-functional theory
within the LDA. A brief preliminary account was presented in
Ref.~\protect\onlinecite{Praha}. To our knowledge, this is the first
study of this type in a ferroelastic temperature-driven system, and
can be considered a benchmark of the capability of these ab-initio
methods to reproduce ferroelastic structural instabilities.  The
case of $\text{CaCl}_2$ is particularly simple: having a rutile-type
structure, it undergoes a tetragonal-orthorhombic transition which is
well characterized as pseudoproper ferroelastic.\cite{CaCl2,Unruh} The
order parameter is one-dimensional and the eigenvector of the unstable
optic mode is fully determined by symmetry arguments. This allows an
investigation of the origin of the structural instability by
considering only a few degrees of freedom of the structure. In
comparison with the case of the ferroelectric perovskites, another
simplifying feature is the fact that the unstable mode is
non-polar, so it is not necessary to include long-range dipolar
interactions in any subsequent modeling of the relevant degrees of
freedom as a local-mode effective hamiltonian.

In this work, the existence of an orthorhombic (slightly distorted
rutile structure) ground-state of the system is explored, and the
total energy is parametrized as a function of the relevant
degrees of freedom. Special emphasis is put on a
comparison of the resulting energy map with a temperature dependent
Landau potential, and the resulting experimental behavior. We also
investigate the importance of other secondary degrees of freedom.

The paper is organized as follows. In Sec. II we describe the
computational details of the first-principles calculations. In
Sec. III we present the ground state of $\text{CaCl}_2$ obtained from
our ab-initio calculations and its comparison with
experiment. Temperature effects are discussed in detail in Sec. IV. In
Sec. V we demonstrate the importance of secondary degrees of freedom
in the determination of the ground state of the system and present
the results obtained from a direct ab-initio minimization of the
orthorhombic structure of $\text{CaCl}_2$. 
The effects of external stresses exerted on CaCl$_2$ are also
considered. Most of the calculations were performed with a
pseudopotential method but the results were cross checked against the
highly accurate all-electron linearized augmented plane wave (LAPW)
method. The latter was mainly used to study the relative energies
between the experimental CaCl$_2$ structure and the denser-packed
fluorite-type structure in both CaCl$_2$ and SrCl$_2$. In Sec. VI we
discuss the competition between the three different structures
(fluorite-type, rutile-type and orthorhombic CaCl$_2$-type). Finally
we draw some relevant conclusions in Sec. VII.

\section{Details of the ab-initio calculations}

Part of the calculations presented in this paper were performed using
density-functional theory within the pseudopotential approach with
plane waves (PW). For these the exchange-correlation energy was
evaluated within the local-density approximation, (LDA)\cite{HK,KS}
using the Perdew and Zunger parametrization\cite{PZ} of the Ceperley
and Alder\cite{CA} interpolation formula for a homogeneous electron
gas.  First-principles norm-conserving pseudopotentials for Ca and Cl
were generated using the scheme proposed by Troullier and
Martins.\cite{TM} The $\text{Ca}$ pseudopotential was calculated with
a nonlinear core correction in the $3d^{0.20} 4s^{0.50} 4p^{0.15}$
non-spin-polarized valence configuration, with
cutoff radii (in bohr) $r_{cd}=1.29$, $r_{cs}=2.66$, and
$r_{cp}=3.09$. The $\text{Cl}$ pseudopotential was calculated in the
$3s^{2.00} 3p^{4.50} 3d^{0.50}$ non-spin-polarized valence
configuration with cutoff radii $r_{cs}=1.72$, $r_{cp}=1.48$, and
$r_{cd}=2.21$. A plane-wave basis set up to a kinetic energy cutoff of
25 ryd and a $4\times4\times5$ Monkhorst-Pack\cite{Monkhorst} shifted
k-point sampling in the Brillouin zone (BZ) were considered. Both the
basis set and k-point sampling were successfully tested to give
converged total energies. A modified Broyden method\cite{Broyden} was
used for the simultaneous relaxation of the ions and the unit cell
while the symmetry was preserved.  Relaxation of the free atomic
coordinates was considered accomplished when atomic forces were less
than 0.05 mhartree/bohr, and the unit-cell relaxation stopped when the
stresses were smaller than 1 MPa.

The full-potential LAPW method as implemented in the WIEN97 program
\cite{wien97} was used to obtain the results presented in Sec. VI. The
``muffin tin'' radius was chosen to be 2.3 bohr for all atoms involved
in the present calculations. The maximum number of plane waves is
determined by the parameter $R_{\rm mt}K_{\rm max}$, which was
8 in our calculations ($R_{\rm mt}$ is the radius of the atomic
sphere and $K_{\rm max}$ is the largest reciprocal wavevector used in
the LAPW basis set). This led to about 1000 plane waves to describe the
valence and semi-core states. Local orbitals were used to treat the Ca
3$s$,3$p$ and Cl 3$s$ semi-core states.  The density of the core
electrons was computed in the crystal potential of each iteration in
the self-consistent cycles (thawed core). Exchange and correlation was
treated within the LDA using the Perdew and Wang
parametrization\cite{PA} of the Ceperley and Alder data.  For
tetragonal and orthorhombic structures we took a $5\times5\times8$
Monkhorst-Pack shifted k-point sampling in the BZ to give converged
total energies, and a $12\times12\times12$ sampling for cubic
structures.  Self-consistency was reached when the total energy of
consecutive iterations changed by less than 0.01 mhartree. The forces
were computed according to Kohler {\textit et al.\/}\cite{kohler} and
relaxation of the free atomic coordinates was completed when the
atomic forces were less than 0.5 mhartree/bohr.

\section{The ferroelastic ground state of $\text{CaCl}_2$}

$\text{CaCl}_2$ undergoes a second order ferroelastic phase transition
at 491K from a high-temperature tetragonal rutile-type structure
(space group $P4_2/mnm, Z=2$) to an orthorhombic one (space group
$Pnnm,Z=2$). The phase transition is therefore equitranslational, with
an order parameter of $B_{1g}$ symmetry (antisymmetric for operations
$4$ and $m_{xy}$). This is in fact the symmetry of the orthorhombic
spontaneous strain $\epsilon=(\epsilon_1-\epsilon_2$)/2.  The
tetragonal high-temperature structure has also a single optical mode
of $B_{1g}$ symmetry, which according to spectroscopic observations
strongly softens as the temperature is lowered and causes the
structural instability.\cite{CaCl2} Thus, the case of $\text{CaCl}_2$
is a textbook example of a pseudoproper ferroelastic transition, where
the actual order parameter $Q$ is the amplitude of an optical mode,
while the spontaneous strain, being of the same symmetry, is
bilinearly coupled to it.

We started the ab-initio analysis by searching for the tetragonal
rutile-type structure of minimal energy, henceforth called the
(ab-initio) parent structure. The atomic positions, constrained within
the symmetry $P4_2/mnm$ (a single free atomic parameter), and the
ratio $c/a$ of the tetragonal unit cell were relaxed for different
unit cell volumes. For each volume the total energy of the relaxed
structure was calculated and the corresponding results were fitted
using the Murnaghan equation of state \cite{Murna} to obtain the
volume of the $P4_2/mnm$ structure of minimal energy. The calculated
parent structure is compared in Table~\ref{table1} with the
experimental high-temperature structure extrapolated at
0~K.\cite{Thesis}
\begin{table}
\caption{\label{table1}
Computed structural parameters and bulk modulus for tetragonal $\text{CaCl}_2$
compared with the experimental values extrapolated to 0~K
(Ref.~\protect\onlinecite{Thesis}). The $x$ atomic coordinate of the
Cl atom at 500~K is given in lattice units. We include both PW and LAPW
results for comparison.}
\begin{ruledtabular}
\begin{tabular}{ldddd}
 &\multicolumn{1}{c}{\text{PW}}
 &\multicolumn{1}{c}{\text{Experiment}}
 &\multicolumn{1}{c}{\text{Diff.(\%)}}
 &\multicolumn{1}{c}{\text{LAPW}} \\
\hline
$a(\text{\AA})$          & 6.302  & 6.343  & 1 & 6.242 \\
$c(\text{\AA})$          & 4.117  & 4.142  & 1 & 4.068  \\
$x(\text{Cl})$           & 0.3028 & 0.3040 & 0.4 & 0.3032 \\
$B_{\rm 0}(\text{GPa})$  & 39     &        &     & 41 \\
\end{tabular}
\end{ruledtabular}
\end{table}
The agreement between PW and experiment is remarkable considering that
LDA is known to underestimate unit cell volumes and thermal effects
influence the experimental high-temperature structure. Also shown in
the table are the structural parameters obtained using the LAPW
method. The deviation between PW and LAPW, both using the
LDA, is most likely caused by the different treatment of the Ca-3$p$
semi-core states: indirectly through the use of non-linear core
corrections in PW, and properly in LAPW.

\begin{figure}
\includegraphics[angle=-90,width=\columnwidth]{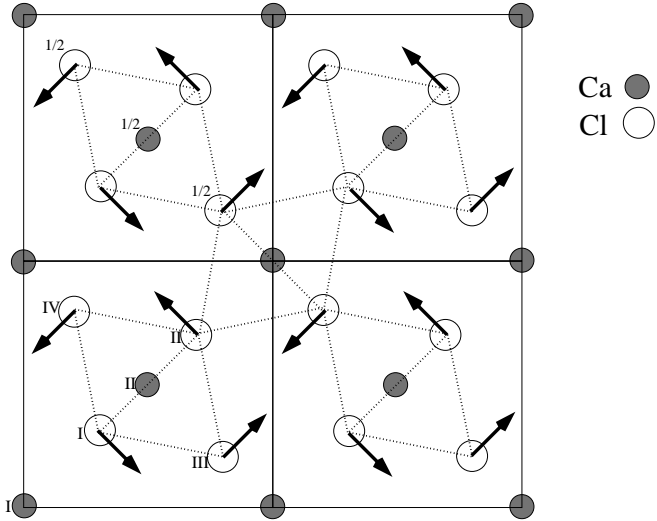}
\caption{\label{fig:sketch}
Sketch showing the parent structure of $\text{CaCl}_2$
and the eigenvector of the unstable $B_{1g}$ mode viewed down the $z$
axis. The mode is close to a rotation around the $z$ axis of the
$\text{CaCl}_6$ octahedra as rigid units.}
\end{figure}

A sketch of the structure is shown in Fig.~\ref{fig:sketch}. Each calcium
cation is surrounded by a distorted octahedron of chlorine anions of
$mmm$ symmetry.  Ideal octahedral coordination would require
$c/a=2-\sqrt{2}=0.5858$ and $x=(2-\sqrt{2})/2=0.2929$ as compared with
values of $0.6533$ and $0.3028$ for our parent structure. This
typical distortion of anion octahedra in rutile type structures has
been explained as a simple mechanism for maximizing the anion-anion
distance, which minimizes its repulsion energy (maximum volume per unit
cell), while keeping fixed the cation-anion distances.\cite{Keeffe}
Indeed, the unit cell volume of the rutile structure would be maximal
under these constraints for $c/a=0.6325$ and $x=0.3000$, \cite{Keeffe}
in fair agreement with the values calculated ab-initio here and those
experimentally observed in $\text{CaCl}_2$ and other rutile
structures, including rather less ionic ones. \cite{bonding}

\begin{table}
\caption{\label{table2}
Calculated frequencies (given in
$\text{cm}^{-1}$) of all zone-center phonons of tetragonal $\text{CaCl}_2$
compared with the available (Raman-active) experimental frequencies
(taken from Ref.~\protect\onlinecite{CaCl2}).}
\begin{ruledtabular}
\begin{tabular}{cddd}
irrep 
 &\multicolumn{1}{c}{\text{Theory}}
 &\multicolumn{1}{c}{\text{Exp.(600K)}}
 &\multicolumn{1}{c}{\text{Diff.(\%)}}\\
\hline
$A_{1g}$   & 229.1            & 203.0   & 12 \\
$A_{2g}$   & 92.7             &         &    \\
$B_{1g}$   & 31.5i            & 17.1    &    \\
$B_{2g}$   & 261.2            & 246.7   & 6  \\
$E_g$      & 169.8            & 150.0   & 13 \\
$A_{2u}$   & 248.7            &         &    \\
$B_{1u}$   & 71.0, 257.5        &         &    \\
$E_u$      & 76.6, 231.6, 271.0 &         &    \\
\end{tabular}
\end{ruledtabular}
\end{table}

The stability of this ab-initio rutile parent structure with respect
to zone-center distortions was then investigated by computing
the normal mode frequencies at the $\Gamma$
point. All $\Gamma$ modes have real frequencies except the mode of
$B_{1g}$ symmetry, confirming the latter's intrinsic instability (see
Table~\ref{table2}).  The frequencies of the stable modes agree fairly
well with experimental frequencies; deviations of $10\%$ are to be
considered normal if it is taken into account that we are comparing 0~K
calculations with experimental values in a structure stabilized at
high temperatures.  The $B_{1g}$ mode has rather low frequencies in
many other rutile-type materials and often exhibits a strong
temperature variation, but in most systems is stable at all
temperatures.\cite{PtO2,MgF2} In some compounds it can be destabilized
via pressure.\cite{Stishovite,NiF2,GeO2} Obviously, in the case of
$\text{CaCl}_2$, the ``soft'' behavior of this mode is so extreme that
it is unstable at low temperatures. Our calculations, which in
principle explore the configuration space at 0~K, agree with this
experimental fact. The tendency of the $B_{1g}$ mode to be rather soft
can be qualitatively explained by its ``quasi rigid-unit''
nature. Indeed, the mode is close to a pure rigid unit rotation of the
columns (parallel to [001]) of distorted chlorine octahedra forming
the structure. Each such column shares corners with four neighboring
chains (see Fig.~\ref{fig:sketch}), so it can be termed a ``quasi
rigid unit mode'' (QRUM).\cite{Rum} This can be easily checked using
the CRUSH package.\cite{Crush1,Crush2} In fact, the $B_{1g}$ mode does
not correspond exactly to a pure rigid unit rotation around the $z$
axis due to the different lengths of the rotation radius of the
chlorine atoms in the $\text{CaCl}_6$ octahedra. Ab-initio studies of
Karki {\textit et al.\/}\cite{stishovite} show an equivalent situation for
the pressure-induced phase transition in stishovite.

Volume effects on the $B_{1g}$ mode frequency were also
calculated. The mode has a negative Gr\"uneisen parameter of around
$-15$, which is in accord with experimental values in other
rutile-type compounds.\cite{TiO2,SnO2,MgF2gr} The resulting decrease
of the mode frequency with pressure is also a typical feature of RUMs
and QRUMs, and is the reason why in other rutile-type compounds like
rutile itself or stishovite the $B_{1g}$ instability can be attained
at high pressures.

\begin{figure}
\includegraphics
[width=\columnwidth]{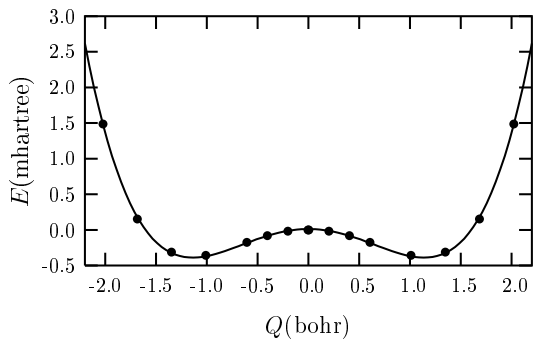}
\caption{\label{fig:wells}
 Total energy (per unit cell) of $\text{CaCl}_2$ as a
function of the amplitude $Q$ of the $B_{1g}$ mode. The solid line
represents the double-well fit indicated in the text.}
\end{figure}

Fig.~\ref{fig:wells} shows the total energy as a function of the
amplitude $Q$ of the mode $B_{1g}$, while maintaining the rest of the
structural parameters at their parent-structure values.
The eigenvector of this mode (see Fig.~\ref{fig:sketch}) has the
form $1/\sqrt{8} (000 ; 000; 1\bar{1}0 ; \bar{1}10 ; 110 ;
\bar{1}\bar{1}0)$, where the displacement vectors are listed in the
order $({\bm e}(\text{Ca}^{\text{I}}),{\bm e}(\text{Ca}^{\text{II}}), 
{\bm e}(\text{Cl}^{\text{I}}),{\bm e}(\text{Cl}^{\text{II}}), 
{\bm e}(\text{Cl}^{\text{III}}),{\bm e}(\text{Cl}^{\text{IV}}))$ 
(atomic labels are indicated in Fig.~\ref{fig:sketch}). The resulting
double-well energy map with a minima of $0.44$ mhartree can be well
fitted using a minimal expansion $E = E_0 + 1/2 \kappa Q^2 + 1/4
\alpha Q^4$, with a mode amplitude of 1.15 bohr at the well minima. The
search for the actual ground state requires, however, an exploration
of the energy of the system as a function of both $Q$ and the
orthorhombic strain, with which it can be bilinearly coupled. In
general, the energy map $E(Q,\epsilon)$ around the parent tetragonal
configuration can be expanded in the form:

\begin{equation}\label{1}
E = E_0 + \frac{1}{2} \kappa Q^2 + \frac{1}{4} \alpha Q^4 +
\frac{1}{2} C_0 \epsilon^2 + \gamma Q \epsilon + \gamma' Q^3 \epsilon
\; .
\end{equation}
 
\begin{figure}
\includegraphics[width=\columnwidth]{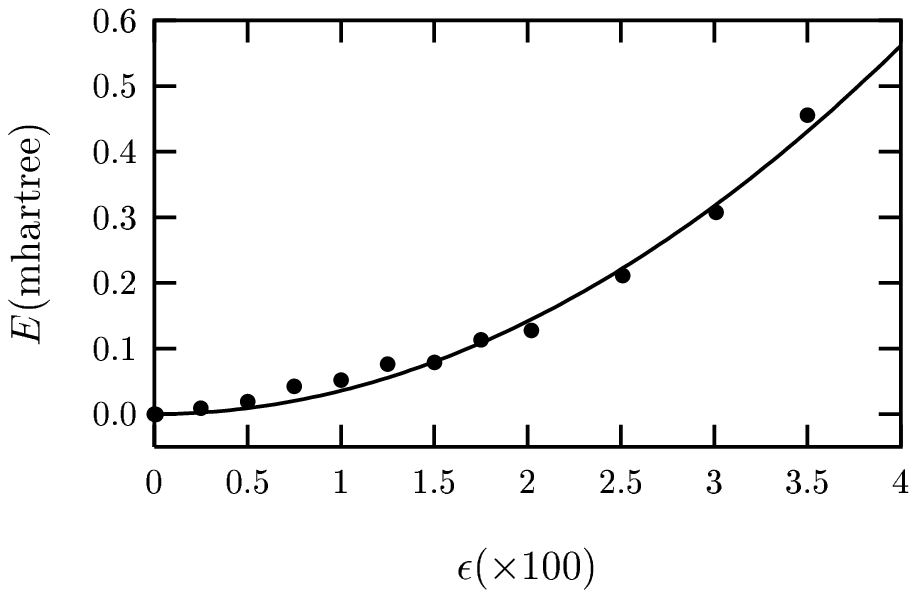}
\includegraphics[width=\columnwidth]{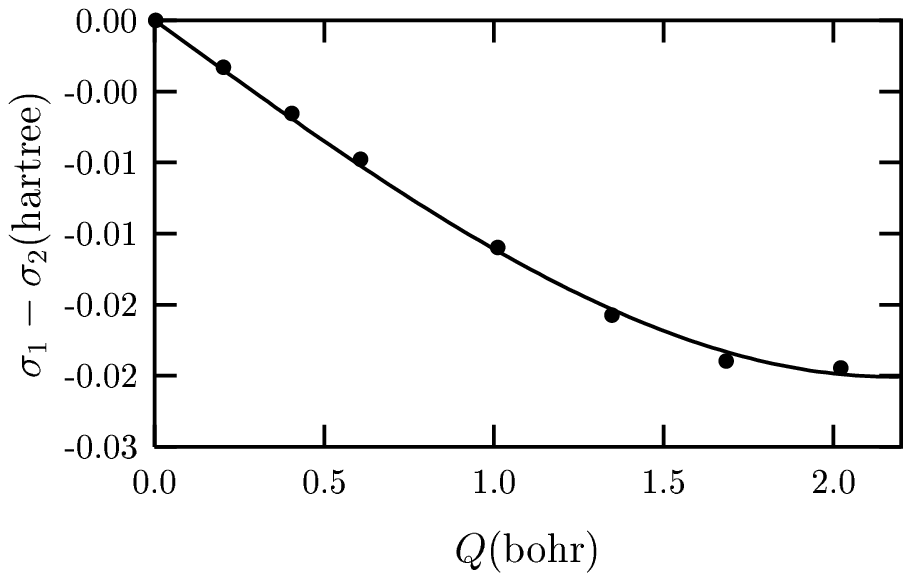}
\caption{\label{fig:fits}
Upper panel: Fit of the calculated total energy for different
strained unit cells to obtain the bare elastic constant $C_0$. Lower panel:
Calculated $\sigma_1-\sigma_2$ stress as a function of the amplitude
$Q$ of the $B_{1g}$ mode at $\epsilon=0$, and the corresponding fit to
$\gamma Q + \gamma' Q^3$.}
\end{figure}

Here $C_0$ is the bare elastic constant of the orthorhombic strain
$\epsilon$ and corresponds to $2(C_{11}-C_{12})$. We have included
only harmonic terms for the elastic energy, but a higher-order
coupling term $Q^3 \epsilon$ is also included as it will be shown
below to be significant. All coefficients in this energy expansion can
be determined from ab-initio calculations: $\kappa$ and $\alpha$ from
the above mentioned fit of the double well of Fig.~\ref{fig:wells} and the
bare elastic constant $C_0$ from the parabolic fit of the total energy
calculated as a function of the strain $\epsilon$ at $Q=0$ (see
Fig.~\ref{fig:fits}a).  The coupling coefficients $\gamma$ and $\gamma'$
can be obtained from the fit of the calculated $\epsilon$-conjugate
stress ${\partial E}/{ \partial \epsilon}$ $(\sigma_{1}-\sigma_{2})$
vs. $Q$ at $\epsilon = 0$ (see Fig.~\ref{fig:fits}b) to the expected
behavior $\gamma Q + \gamma' Q^3$. In Fig.~\ref{fig:fits}b, deviations
from the linear behavior are clearly observed, so the higher
non-linear coupling $Q^3 \epsilon$ is quite significant.
Table \ref{table3} lists the values obtained for these various
coefficients. The resulting energy map $E(Q,\epsilon)$, has three
stationary points: the saddle point at $Q_0=0$, $\epsilon_0=0$,
corresponds to the unstable tetragonal structure, and the other two
symmetry-equivalent stationary points locate the ground state of the
structure at the twin related absolute minima $(Q_m,\epsilon_m)=\pm
(1.27\text{ bohr},0.029)$. The depth of the minima with respect to the
tetragonal parent structure is $0.70$ mhartree. When compared
with the relative minimum along $Q$, it means that nearly $50\%$ of
the energy minimization corresponds to strain relaxation. This should
be compared with the experimental $B_{1g}$ distortion extrapolated to
0~K.  According to Ref.~\protect\onlinecite{Thesis}, the strain
$\epsilon$ at room temperature is 0.014, and the experimental curve
$\epsilon(T)$ can be extrapolated to $\epsilon(0)=0.022$. Assuming an
analogous behavior for $Q(T)$, as expected in pseudoproper
ferroelastics, the experimental value of 0.85 bohr for $Q$ at room
temperature \cite{Thesis} can be extrapolated to 1.33 bohr at 0~K. This
value is in excellent agreement with the ab-initio value while, on the
other hand, the orthorhombic ab-initio strain is about $30\%$
overestimated.

\begin{table}
\caption{\label{table3}
Parameters of the total energy expansions
$E(Q,\epsilon)$ given in Eq.~(\ref{1}) and Eq.~(\ref{2}).}
\begin{ruledtabular}
\begin{tabular}{cdlcdl}
\multicolumn{6}{c}{$B_{1g}$-mode}\\
$\kappa$ &-1.33 &$\text{mh}/\text{bohr}^2$ 
& $\alpha$ & 1.01 &$\text{mh}/\text{bohr}^4$\\ 
\hline 
\multicolumn{6}{c}{$A_{1g}$-mode}\\
$\delta$ & 70.42 &$\text{mh}/\text{bohr}^2$ &\\ 
\hline 
\multicolumn{6}{c}{Couplings}\\
$\gamma$ & -16.6 &$\text{mh}/\text{bohr}$ & $\gamma'$ 
& 0.80 &$\text{mh}/\text{bohr}^3$ \\ 
$a$ & 50 &$\text{mh}/\text{bohr}$ 
& $b$ & -194 &$\text{mh}/\text{bohr}$ \\ 
$a'$ & -1.07 &$\text{mh}/\text{bohr}^3$ & $b'$ 
& 32.1 &$\text{mh}/\text{bohr}^2$ \\ 
$c'$ & 16.8 &$\text{mh}/\text{bohr}^2$ & & \\ 
\hline 
\multicolumn{6}{c}{Elastic}\\
$C_0$ & 0.68 &$\text{h (18GPa)}$ & $C_+$ 
& 6.95 &$\text{h (184GPa)}$ \\ 
$C_3$ & 3.15 &$\text{h (83GPa)}$ & $C_{+3}$ & 2.15 &$\text{h (57GPa)}$ \\ 
\end{tabular}
\end{ruledtabular}
\end{table}

\section{Temperature effects}

The calculated energy map $E(Q,\epsilon)$ can be used as the starting
point for the ``prediction" of thermal behavior, including the phase
transition.  This can be done, for instance, through the construction
of an effective hamiltonian consistent with the energetics described
by $E(Q,\epsilon)$, and its further analysis through Monte Carlo
simulations.\cite{BaTiO3} As a zeroth-order approximation, however, we
can already perform some sensible comparisons with experimental results, if
we note that the energy map $E(Q,\epsilon)$ can be taken as an
approximation for the free-energy (for given $Q$ and $\epsilon$, in
the sense of a Landau potential) at 0~K, and we make the (rather
drastic) assumption that the temperature effects are restricted to
the thermal renormalization up to positive values of the quadratic
$\kappa$ coefficient, so that it follows a Landau type law $ \kappa
\propto (T-T_0)$. In other words, the coefficients in Table
\ref{table3} can be considered an estimation of the coefficients in
the Landau free energy, except for $\kappa$. For instance, according
to Eq.~(\ref{1}), the actual observable elastic constant corresponding
to $\epsilon$ should soften according to the law:
$C(T)=C_0-\gamma^2/\kappa(T)$, so that the phase transition takes
place when $C(T)$ becomes zero at
$\kappa(T_c)=\gamma^2/C_0$. According to the parameter values of Table
\ref{table3}, this would correspond to a value of $\kappa(T_c)=0.41
\text{ mhartree}/\text{bohr}^2$ which, taking into account the
effective mass of the mode ($m_{\text{Cl}}$ in our $Q$ units) represents a
mode frequency $\omega_{B_{1g}}=17~\text{cm}^{-1}$, in good agreement with
the experimental value \cite{CaCl2} of $14~\text{cm}^{-1}$. 

\begin{figure}
\includegraphics[width=\columnwidth]{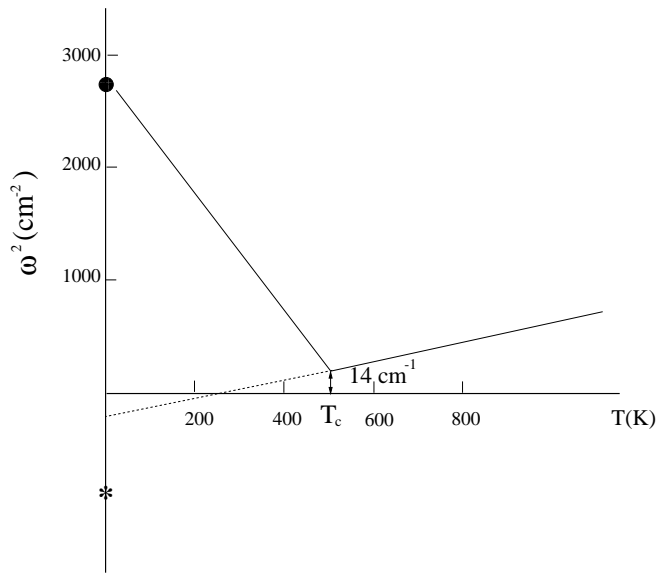}
\caption{\label{fig:omega}
Sketch of the experimental temperature behavior of the square
of the frequency of the soft mode in $\text{CaCl}_2$
(Ref.~\protect\onlinecite{CaCl2}).  The asterisk and the black circle
at 0~K represent ab-initio values of $\omega^2$ for $Q$=0, $\epsilon$=0
and at the absolute minima of the total-energy map, respectively.}
\end{figure}

An expansion of the free energy including a higher-order coupling
term $Q^3\epsilon$ as that given in Eq.~(\ref{1}) for $E(Q,\epsilon)$,
was proposed to study the temperature behavior of the soft optic
phonon observed by Raman spectroscopy experiments. \cite{CaCl2} In
particular, this coupling term was included to explain the large value
of the slope ratio $r$ of $\omega^2(T)$ below and above the phase
transition, as compared to the value of $-2$ expected from the
simplest Landau expansion. By using this extended free-energy
expansion, it was demonstrated that the slope ratio of $\omega^2(T)$
is no longer $-2$ (the mean-field value), but a function which depends
on the free-energy expansion coefficients: 
$r= (k-4\alpha)/2(\alpha-k)$, where $k=4\gamma\gamma'/C_0$.
\cite{CaCl2} Using this result, the calculated non-linear coupling
coefficient $\gamma'$ predicts a slope ratio of $\omega^2(T)$ of
$-1.9$, which is not only quite far from the observed $r$ of $-6.5$,
but it corrects the classical value $-2$ in the opposite
direction. Trying to explain such a slope ratio through this
non-linear term in the free energy would require a coefficient
$\gamma'$ one order of magnitude larger than the ab-initio calculated
value and of opposite sign, which seems difficult to accept even
considering any reasonable thermal renormalization effect. Therefore,
our results indicate that this non-classical slope ratio must have a
quite different origin. In this respect, it is quite interesting to
compare the values of $\omega^2$ at 0~K extrapolated from the
experimental linear behavior of $\omega^2(T)$, below $T_c$ (see
Fig.~\ref{fig:omega}), with the theoretical expected value obtained
from the curvature along $Q$ at the absolute minima of the ab-initio
energy map. The experimental value is $3.74~\text{
mhartree}/\text{bohr}^2$ ($2790~\text{cm}^{-2}$) and the theoretical
one $\kappa + 3 \alpha Q_m^2 + 6 \gamma' \epsilon_m Q_m= 3.73 \text{
mhartree}/\text{bohr}^2$ ($2780~\text{cm}^{-2}$), a
surprising excellent agreement.
This suggests that the strong slope
change of the linear temperature behavior of the mode frequency has a
rather simple origin: the $Q$-stiffness coefficient approaches
linearly the ground state value once the system is below $T_c$. This
implies, in general, a correction to the usual Landau description
through a change of slope of the linear decrease of the quadratic
order parameter coefficient around $T_c$. This kind of behavior has
been actually observed in Monte Carlo simulations of the $\Phi^4$
model when described by means of an effective Landau potential.
\cite{Phi4,svet,janssen} The curvature along $Q$ at the tetragonal
maximum is $-990~\text{cm}^{-2}$, indicated in Fig.~\ref{fig:omega}
with an asterisk. This is four times lower than the extrapolated value
from the experimental linear behavior of $\omega^2(T)$ above
$T_c$. The deviation of this extrapolated value from the actual 0~K
value is also normal in simplified models like the $\Phi^4$, and is
related to the displacive degree of the transition. Only in the
displacive limit would one expect both values to
agree.\cite{svet,manu} A deviation factor of four would correspond to
a quite normal and realistic intermediate regime where couplings and
ground state energy scales are of the same order of magnitude.

\section{Secondary degrees of freedom}

Until now, in our analysis of the energy we have only considered the
primary degrees of freedom which determine the symmetry of the crystal
below the transition, i.e., we have expanded the energy only in terms
of the amplitude $Q$ of the optic mode $B_{1g}$ and the orthorhombic
strain $\epsilon$. However, we must also consider other secondary
degrees of freedom which do not determine the change of symmetry but
are compatible with it and may be important in the study of the ground
state of $\text{CaCl}_2$. In our particular case, these are easily
identified as the amplitude $R$ of the optic mode $A_{1g}$ and the
tetragonal strain components $\epsilon_+=(\epsilon_1+\epsilon_2)/2$
and $\epsilon_3$. The eigenvector of the mode $A_{1g}$ has the form
$1/\sqrt{8} (000 ; 000; 110 ; \bar{1}\bar{1}0 ; 1\bar{1}0 ;
\bar{1}10)$ (in the same basis used above for the $B_{1g}$
mode). Using these new variables, we can write down a new expansion
for $E$:

\begin{eqnarray}\label{2}
E = E_0 + \frac{1}{2} \kappa Q^2 + \frac{1}{4} \alpha Q^4 + 
\frac{1}{2} C_0 \epsilon^2 + \gamma Q \epsilon + \gamma' Q^3 \epsilon + 
\nonumber\\
+ \frac{1}{2} \delta R^2 + \frac{1}{2} C_+ \epsilon_+^2 + 
\frac{1}{2} C_3 \epsilon_3^2 + C_{+3} \epsilon_+ \epsilon_3 + 
\nonumber\\
+ a R \epsilon_+ + b R \epsilon_3 +
+ a' R Q^2 + b' \epsilon_+ Q^2 + c' \epsilon_3 Q^2 
\;\;.
\end{eqnarray}

Our calculations show that allowed fourth order terms such as $R^4,
\epsilon^4, \epsilon_+^4, \epsilon_3^4, Q\epsilon^3$ are not
significant and thus have not been included in the previous
expansion. We have also checked that the contribution of other
higher-order coupling terms is several orders of magnitude smaller
than those in Eq.~(\ref{2}).  $C_+$, $C_3$ and $C_{+3}$ are bare
elastic constants and correspond to $2(C_{11}+C_{12})$, $C_{33}$ and
$2C_{13}$, respectively. All the coefficients in the expansion given
by Eq.~(\ref{2}) have been determined from ab-initio
calculations. $\delta$ comes directly from the phonon frequency given
in Table~\ref{table2}. The bare elastic constants $C_+$ and $C_3$ were
obtained from the parabolic fit of the total energy for $Q=0$ versus
$\epsilon_+$ and $\epsilon_3$ at $Q=0$ respectively. The coefficients
$a$ and $b$ result from the calculation of the R-conjugate force $F_R$
as a function of $\epsilon_+$ and $\epsilon_3$ respectively, and the
elastic constant $C_{+3}$ from the calculation of $\sigma_+$
($\sigma_1+\sigma_2$) versus $\epsilon_3$. The coupling coefficient
$a'$ is obtained from the parabolic fit of the force $F_{R}$ as a
function of the amplitude $Q$ of the mode $B_{1g}$. $b'$ and $c'$ are
obtained by fitting $\sigma_+$ and $\sigma_3$ versus the amplitude $Q$
to a parabola. Table~\ref{table3} gives a full list of the values
obtained for these coefficients.

The resulting energy map can be analyzed in a way similar to that used in the
case of Eq.~(\ref{1}). This is evident if we notice that the energy
$E$ in Eq.~(\ref{2}) can be rewritten in an equivalent form to
that given in Eq.~(\ref{1}) but with the quartic coefficient $\alpha$
renormalized to a smaller value $\alpha'$ given by $0.656\text{
mhartree}/\text{bohr}^4$ when one introduces the minimum conditions
$(\partial{E}/\partial{R})_0=0$,
$(\partial{E}/\partial{\epsilon_+})_0=0$ and
$(\partial{E}/\partial{\epsilon_3})_0=0$ into Eq.~(\ref{2}). The most
important contribution to the renormalization of $\alpha$ comes from
the $\epsilon_+Q^2$ coupling term, with almost $90\%$ of the total
value.  The energy map has three stationary points: the trivial one
$Q_0=0, \epsilon_0=0$ with $E=0$ which corresponds to a maximum, and
the ground state $(Q_m,\epsilon_m)=\pm(1.54\text{ bohr},0.033)$ with
$R=-0.01\text{ bohr},\epsilon_+=-0.010,\epsilon_3=-0.004$ and the
depth of the minima $\Delta E=-0.98\text{ mhartree}$. As expected,
the minima are deeper because of a smaller value of the
effective quartic coefficient.  Furthermore, the ground state values
of $Q$ and $\epsilon$ have increased, worsening the agreement obtained
with experiment when secondary modes were neglected. In fact, the
error in the unit cell volume due to the LDA over-binding might be
crucial when one considers this issue. 
The values of $\epsilon_+$ and $\epsilon_3$ can be compared to the
experimental ones~\cite{Unruh} extrapolated to 0K ($\epsilon_+=-0.006$
and $\epsilon_3=-0.001$). 

The deviation from the classical value of the ratio of the slopes of
the soft-mode temperature behavior could in principle be related to
the secondary strain degrees of freedom analyzed here. As shown above
in general, these additional variables renormalize the quartic
coefficient $\alpha$ to a lower value $\alpha'$. When considering
secondary spontaneous strains clamped at optical frequencies the slope
ratio becomes $r=-2\alpha/\alpha'$. In the present case, for the
calculated strain coupling strengths, this could mean a correction of
$r$ to about $-3$, still very far from the
$-6.5$ observed. On the other hand, these secondary modes or degrees
of freedom also affect the calculation done above for the curvature
along $Q$ of the ab-initio energy map at the absolute minima and the
corresponding ab-initio ground state soft-mode frequency. This becomes
5.34 mhartree/bohr$^2$ (3980 cm$^{-1}$), a huge increase of about 40\%
with respect to the value obtained disregarding secondary modes (see
Fig.~\ref{fig:omega}). This is significantly larger than the
extrapolated experimental 0K value. Therefore one can conclude that
the comparison with experiment worsens systematically when secondary
strain deformations directly related to changes in the unit cell
volume are included.

\begin{table}
\caption{\label{table4}
Structural parameters and bulk modulus of the
fully relaxed orthorhombic phase of $\text{CaCl}_2$ compared with the
experimental values extrapolated to 0~K
(Ref.~\protect\onlinecite{Thesis}). The $x$ and $y$ atomic coordinates
for Cl atoms at room temperature are given in lattice units.}
\begin{ruledtabular}
\begin{tabular}{l|ddd}
 &\multicolumn{1}{c}{\text{Computed}}
 &\multicolumn{1}{c}{\text{Experiment}}
 &\multicolumn{1}{c}{\text{Difference(\%)}}\\
\hline
$a(\text{\AA})$  & 6.429   & 6.446  & 1 \\
$b(\text{\AA})$  & 6.054   & 6.167  & 2 \\
$c(\text{\AA})$  & 4.088   & 4.137  & 1 \\
$x(\text{Cl})$   & 0.347  & 0.325 & 7 \\
$y(\text{Cl})$   & 0.255  & 0.275 & 7 \\
$B_0 (\text{GPa})$  & 31    & & \\
\end{tabular}
\end{ruledtabular}
\end{table}

The results obtained above can be compared with the outcome of a
direct search for the orthorhombic ground state of $\text{CaCl}_2$.
This comparison also leads to an estimation of the error in the
calculations. We minimized the total energy of an orthorhombic unit
cell with respect to the volume, allowing for the relaxation of the
ratios $b/a$ and $c/a$ and the atomic positions, constrained within
the $Pnnm$ space group (two free atomic parameters).
The calculated structure is compared in
table~\ref{table4} with the experimental orthorhombic structure
extrapolated to 0~K. The agreement is very good considering the well
known LDA underestimation (the predicted volume of the orthorhombic 
unit cell is $V_{\text ort}=1075$~bohr$^3$, to be compared with the value
extrapolated to 0K from experiment (1111~bohr$^3$)). Using the
calculated structural parameters listed in tables \ref{table1} and
\ref{table4} we see that in the absolute orthorhombic minima
$\epsilon=0.030$, $\epsilon_+=-0.010$ and $\epsilon_3=-0.007$, which
agrees fairly well with the calculated strains from
Eq.~(\ref{2}). From the calculated atomic positions of the chlorine
atoms we also obtain the amplitude of the $B_{1g}$ mode, $Q=1.55$ bohr
and the $A_{1g}$ mode, $R=-0.06$ bohr. The energy difference between
the orthorhombic and tetragonal structures is $-0.92$ mhartree, to be
compared with the depth of the minima, $\Delta E = -0.98$ mhartree,
from Eq.~(\ref{2}). Hence, these results are consistent with the
ground state that we have found considering the energy expansion given
by Eq.~(\ref{2}) with errors of the order of $10 \%$.

\section{Energy-volume curves and relative stability of different
 structures}
\label{sec:compet}

We have found it useful to look at the issue of the ferroelastic
instability in $\text{CaCl}_2$ from the point of view of the analysis
of energy-volume curves, traditional in the field of first-principles
calculations. Apart from the consideration of the two phases involved
in the transition, a question which comes to mind is whether there
might be other structures that could compete for stability with the
experimentally observed $\text{CaCl}_2$ structure. A good candidate is
the cubic fluorite structure, with the anions placed in a tetrahedral
environment and an eightfold coordination of the cations, which is the
stable arrangement for the chemically similar compound
$\text{SrCl}_2$.

\begin{figure}
\includegraphics[angle=-90,width=\columnwidth]{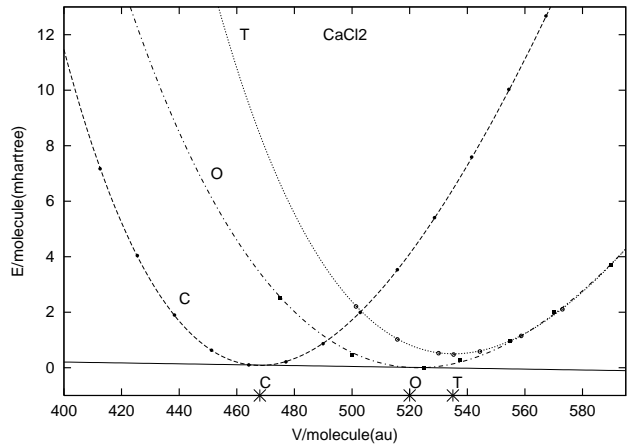}
\caption{\label{fig:e_v_cacl2} 
Total-energy curves for $\text{CaCl}_2$
as a function of the unit cell volume per formula unit (f.u.) for the
cubic fluorite-type (C), tetragonal rutile-type (T) and orthorhombic
$\text{CaCl}_2$-type (O) structures.  Asterisks mark the volumes per
f.u. at the minima. The common tangent of the O and C curves is shown
to highlight the very small energy difference between their minima.}
\end{figure}

The calculations in this section are carried out using the the highly
accurate all-electron LAPW method, which for the structural properties
of $\text{CaCl}_2$ yields essentially the same results as the PW code,
as has been shown in Tab.~\ref{table1}, but is in principle more
suitable to analyze small energy differences between structures.  We
have computed the energy-volume curves for $\text{CaCl}_2$ in the
rutile-type (T), the $\text{CaCl}_2$-type orthorhombic (O), and the
fluorite-type cubic (C) structures, and show them in
Fig.~\ref{fig:e_v_cacl2}. The asterisks under C, O, and T mark the
equilibrium volumes per formula unit (468, 520, and 535 bohr$^3$,
respectively).

\begin{figure}
\includegraphics[angle=-90,width=\columnwidth]{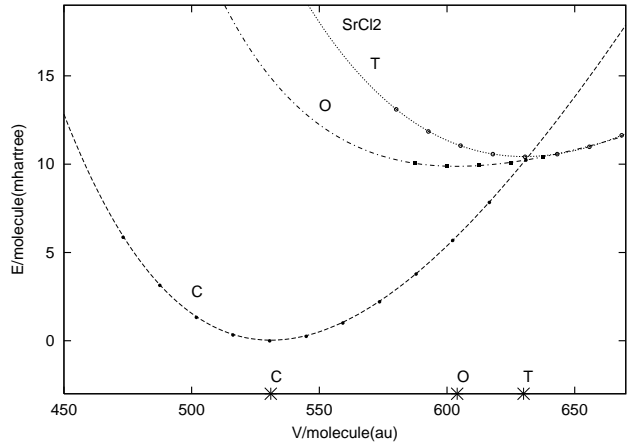}
\caption{\label{fig:e_v_srcl2}
Total-energy curves for $\text{SrCl}_2$
as a function of the unit cell volume per formula unit (f.u.) for the
cubic fluorite-type (C), tetragonal rutile-type (T) and orthorhombic
$\text{CaCl}_2$-type (O) structures.  Asterisks mark the volumes per
f.u. at the minima.}
\end{figure}

To put the case of $\text{CaCl}_2$ in perspective, we have performed a
similar analysis for $\text{SrCl}_2$. Fig.~\ref{fig:e_v_srcl2} shows
the total-energy curves for the same three structure-types as for
$\text{CaCl}_2$. The minimum of the total-energy curve for the
fluorite-type cubic structure is located at 531 bohr$^3$, to be
compared with the observed experimental value of 574 bohr$^3$. The
difference can be attributed to the use of the LDA.

The $\text{CaCl}_2$-type orthorhombic structure is very similar to the
rutile-type tetragonal structure. So, it was to be expected that the
energy difference is small both in $\text{CaCl}_2$ and in
$\text{SrCl}_2$ (0.48 and 0.54 mhartree/f.u., respectively). In both
cases the energies continuously merge as the volume increases, since
the $B_{1g}$ optic mode becomes stable and the distortions
disappear. This is the signature of a group-subgroup symmetry-breaking
transition in an energy-volume diagram. SrCl$_2$ in the rutile-type
structure would have an unstable $B_{1g}$ mode similar to CaCl$_2$,
including the negative Gr\"uneisen parameter.

In $\text{CaCl}_2$, the three structures have very similar energies.
The fluorite-type cubic structure is more stable than the rutile-type
tetragonal structure (the energy difference is 0.38
mhartree/f.u.). So, the cubic and tetragonal structures are
competitive and only the orthorhombic distortion reduces the energy by
an additional 0.1 mhartree/f.u. below the cubic one. As shown in
Fig.~\ref{fig:e_v_cacl2}, a very small pressure would be enough to
transform $\text{CaCl}_2$ into the denser fluorite-type cubic
structure. This small energy difference has led us to re-do the
calculation with the generalized gradient approximation (GGA) for
exchange and correlation.\cite{pbe} In this case the fluorite-type
structure is about 8 mhartree/f.u. higher than the tetragonal one,
significantly increasing the critical pressure.  This finding is
consistent with the work of Zupan {\textit et al.\/}\cite{Zupan} on
the stishovite/alpha-quartz phase transition in SiO$_2$, which showed
that the compressed phase is too stable within the LDA, whereas the GGA
corrects for this over-binding of the LDA and thus agrees better with
experiment. In that paper it was argued that it is a general feature
of the LDA to favor more compact (isotropic) structures, while
the GGA enhances charge inhomogeneities and thus stabilizes the
higher-volume and lower-symmetry phase. The present case is another
example of this general behavior.

In SrCl$_2$, the ground state of the system corresponds to the
fluorite-type cubic structure, in accord with the experimental
observation. The energy difference with respect to the orthorhombic
structure (within the LDA) is 9.88 mhartree/f.u., much larger than in
the case of CaCl$_2$. This can be understood by considering the
non-bonding forces between the cations. The corresponding potential
grows steeply at short distances due to strong repulsive forces. For
this reason the relative energies are sensitive to the first-neighbor
Ca-Ca and Sr-Sr distances. In SrCl$_2$ the cations in the tetragonal
structure are very close to each other and thus raise the total
energy. In CaCl$_2$, however, the cations are not so close and thus
the corresponding energy difference between the structures is
relatively small.

\section{Conclusions}

Ab-initio calculations within the LDA approximation without volume
corrections are sufficient to reproduce the main features of the
ferroelastic instability in $\text{CaCl}_2$. Our analysis of the
energetics of the crystal confirm a pseudoproper ferroelastic
mechanism with an unstable optic mode as the microscopic origin of the
experimental phase transition. The elastic instability arises due to
the coupling between this soft $B_{1g}$ optic mode and the
orthorhombic spontaneous strain $\epsilon$. The existence of QRUMs in
the rutile-type framework structure has been shown to play an
essential role as well. The calculated orthorhombic ground state
agrees fairly well with the experimental low temperature structure. We
have identified and calculated the anharmonic terms which have
relevance in the determination of the ground state.  The secondary
degrees of freedom have been also considered.  Our energy-map
calculations rule out the relevance of higher anharmonic couplings
proposed in previous literature to explain the ratio of the slopes of
the squared frequencies. Although the system modifies its ground state
noticeably when volume/strain corrections are considered, the LDA
approximation has been sufficient for obtaining a realistic picture of
the microscopic mechanism of this temperature driven ferroelastic
system.

\begin{acknowledgments}

This work was supported in part by the UPV research grants
060.310-EA149/95 and 063.310-G19/98, and by the Spanish Ministry of
Education grant PB98-0244. We gratefully acknowledge valuable
discussions with E.K.H. Salje and K. Parlinski and helpful comments
from J. \'I\~niguez and S. Ivantchev. We are also indebted to
M. T. Dove for his help with the program CRUSH. J.A.V. was supported
by the Spanish Ministry of Education and appreciates the warm
hospitality at the TU in Vienna during the realization of part of this
work.

\end{acknowledgments}

\end{document}